\documentclass[twocolumn]{article}
\usepackage[a4paper, total={6in, 9in}]{geometry}
\usepackage{graphicx} 
\usepackage[square,numbers]{natbib}  
\usepackage{multirow} 
\usepackage{booktabs} 
\usepackage{tikz}
\usepackage{amssymb}  
\usepackage{amsmath}
\usepackage{soul}
\usepackage{caption}
\usepackage{url}
\usepackage{authblk}
\usepackage{pgfplots}
\usepackage{algorithm}
\usepackage{algpseudocode}
\pgfplotsset{compat=1.16}

\usepackage[capitalise]{cleveref}

\title{Interactive 3D Segmentation for Primary Gross Tumor Volume in Oropharyngeal Cancer}
\author[1]{Mikko Saukkoriipi}
\author[1]{Jaakko Sahlsten}
\author[1]{Joel Jaskari}

\author[2]{Lotta Orsmaa}
\author[2]{Jari Kangas}
\author[2]{Nastaran Rasouli}
\author[2]{Roope Raisamo}

\author[3]{Jussi Hirvonen}
\author[3]{Helena Mehtonen}
\author[3]{Jorma Järnstedt}

\author[4]{Antti Mäkitie}

\author[5]{Mohamed Naser}
\author[5]{Clifton Fuller}
\author[6,7]{Benjamin Kann}

\author[1,8*]{Kimmo Kaski}

\affil[1]{Department of Computer Science, Aalto University School of Science, Espoo, Finland}
\affil[2]{Faculty of Information Technology and Communication Sciences, Computing Sciences, University of Tampere, Tampere, Finland}
\affil[3]{Department of Radiology, Tampere University, Faculty of Medicine and Health Technology, and Tampere University Hospital, Tampere, Finland}
\affil[4]{Department of Otorhinolaryngology–Head and Neck Surgery, Research Program in Systems Oncology, Faculty of Medicine, University of Helsinki and Helsinki University Hospital, Helsinki, Finland}
\affil[5]{Department of Radiation Oncology, The University of Texas MD Anderson Cancer Center, Houston, TX USA}
\affil[6]{Artificial Intelligence in Medicine Program, Mass General Brigham, Harvard Medical School, Boston, MA, USA}
\affil[7]{Department of Radiation Oncology, Dana-Farber Cancer Institute and Brigham and Women’s Hospital, Harvard Medical School, Boston, MA, USA}
\affil[8]{The Alan Turing Institute, British Library, 96 Euston Rd, London NW1 2DB, United Kingdom}
\affil[*]{Corresponding author}

\date{\today}
\begin{document}

\twocolumn[
\begin{@twocolumnfalse}
\maketitle
\begin{abstract}


\noindent
The main treatment modality for oropharyngeal cancer (OPC) is radiotherapy, where accurate segmentation of the primary gross tumor volume (GTVp) is essential. However, accurate GTVp segmentation is challenging due to significant interobserver variability and the time-consuming nature of manual annotation, while fully automated methods can occasionally fail. An interactive deep learning (DL) model offers the advantage of automatic high-performance segmentation with the flexibility for user correction when necessary. In this study, we examine interactive DL for GTVp segmentation in OPC. We implement state-of-the-art algorithms and propose a novel two-stage Interactive Click Refinement (2S-ICR) framework. Using the 2021 HEad and neCK TumOR (HECKTOR) dataset for development and an external dataset from The University of Texas MD Anderson Cancer Center for evaluation, the 2S-ICR framework achieves a Dice similarity coefficient of 0.713 ± 0.152 without user interaction and 0.824 ± 0.099 after five interactions, outperforming existing methods in both cases.

\end{abstract}
\end{@twocolumnfalse}
]


\clearpage 
\section{Introduction}



Oropharyngeal cancer (OPC) is a subtype of head and neck squamous cell carcinoma that predominantly affects the the tonsils and the tongue base and poses substantial challenges in medical imaging and treatment. Early detection and effective management of OPC are critical for enhancing the patient outcomes including quality of life and survival~\cite{nunez2024opcimpact}. Magnetic Resonance Imaging (MRI), Computed Tomography (CT), and Positron Emission Tomography (PET) are the primary modalities used for the initial staging, and radiotherapy (RT) planning, and follow-up of OPC~\cite{andrearczyk_hecktor2021_overview}. RT is a pivotal treatment modality for OPC, but it relies on the labor-intensive and error-prone manual or semi-automatic segmentation of the primary gross tumor volume (GTVp)~\cite{andrearczyk_hecktor2021_overview}. An accurate segmentation of GTVp in the oropharynx region is particularly challenging due to significant interobserver variability~\cite{rasch2005target, cardenas2022comprehensive, lin2023pluribus}. This challenge not only compromises the efficacy of treatment, but it also increases both the duration and cost of care~\cite{andrearczyk_hecktor2021_overview}. Consequently, there is a need for the development of precise, fast, and cost-efficient automatic segmentation techniques for OPC GTVp that would enhance treatment outcomes and operational efficiency.


An automated segmentation of the OPC GTVp using deep learning (DL) methods has shown good promise in reducing variability and enhancing the precision and reliability of radiotherapy planning~\cite{Iantsen_2021_hecktor_2020_winner, sahlsten2024hecktor, myronenko_hecktor_winner_2022}. However, the segmentation can fall short of the required performance and would necessitate further manual refinement or complete rework by clinicians. In such cases, interactive deep learning presents a compelling approach by facilitating efficient interface for segmentation correction and enhancing clinicians trust in the segmentation model.

A common approach for interactive DL segmentation is a click-based interaction, where the user provides feedback by clicking on coordinates where the DL segmentation requires correction~\cite{wang2022medical_survey}. The click-based interaction can be facilitated by a graphical user interface that allows the user to review the current segmentation, input corrections, and receive immediate feedback from the segmentation model. To our knowledge, the only work that has considered interactive DL for OPC tumor segmentation is~\cite{wei2023opc_slice_idl}, which utilized a slice-based interaction method, where the user manually segments an entire slice of the tumor volume. This approach requires time-consuming re-training of the DL model after each interaction, limiting its practical use in clinical settings. Consequently, there is limited exploration of fast click-based interactive DL segmentation for improving OPC GTVp segmentation, despite its potential benefits. 


DeepGrow~\cite{deepgrow_2019} and DeepEdit~\cite{DeepEdit_2022}, both integrated within the Medical Open Network for AI (MONAI)~\cite{cardoso2022monaiopensourceframeworkdeep}, are two click-based interactive deep learning tools tailored for segmentation tasks. While both tools facilitate user guided segmentation, DeepGrow mandates that at least one user click needs to be provided in order to segment an image. In contrast, DeepEdit enhances this capability by also supporting segmentation without user input. To function in both automatic and interactive modes, DeepEdit employs a hyperparameter known as "click-free iterations," which sets the likelihood of non-interactive training iterations. However, this configuration introduces a trade-off between non-interactive and interactive segmentation performance. Evaluations on multiple 3-dimensional (3D) medical image datasets have shown that the performance of DeepEdit is inferior to traditional DL segmentation methods in non-interactive mode, and inferior to DeepGrow in interactive mode~\cite{DeepEdit_2022}. This raises the question if 3D medical image segmentation can be performed accurately by an interactive DL model in both interactive and non-interactive settings.


In this study, we introduce a novel two-stage Interactive Click Refinement (2S-ICR) framework that is designed to enhance OPC GTVp segmentation performance through manual interactions, while maintaining a high baseline accuracy without user input. Our contributions are as follows. For the first time, we demonstrate the effectiveness of interactive OPC GTVp segmentation from PET-CT scans in volumetric space. We also provide a systematic comparison of two click-based state-of-the-art interactive DL methods alongside the 2S-ICR framework for OPC GTVp segmentation, showing that our method performs comparably or better in both non-interactive and interactive scenarios, effectively setting a new state-of-the-art for this task. Additionally, we show that an ensemble approach can further enhance the effectiveness of our interactive method.

\clearpage
\section{Results}

\subsection{Experimental setup}
To develop and validate our models, we performed 5-fold cross validation with the 2021 HEad and neCK TumOR (HECKTOR) training dataset~\cite{andrearczyk_hecktor2021_overview}. In model training and evaluation, we simulated user interactions with an algorithm similar to~\cite{deepgrow_2019, DeepEdit_2022}. To reduce the variability in results due to the probabilistic nature of simulated interaction events, we evaluated the 5-fold validation performance three times with different random seeds. In addition, we evaluated the out-of-distribution performance of the models with an external validation dataset from The University of Texas MD Anderson Cancer Center (MDA), for which we similarly repeated the validation three times. The DeepEdit method was trained with two different click-free iteration proportions, namely those of 25\% and 50\%, which we denote DeepEdit-25 and DeepEdit-50, respectively. 

\subsection{Interactive segmentation performance}

\begin{figure*}[ht]
  \centering
  \hspace*{-0.06\linewidth}
  \includegraphics[width=1.1\linewidth]{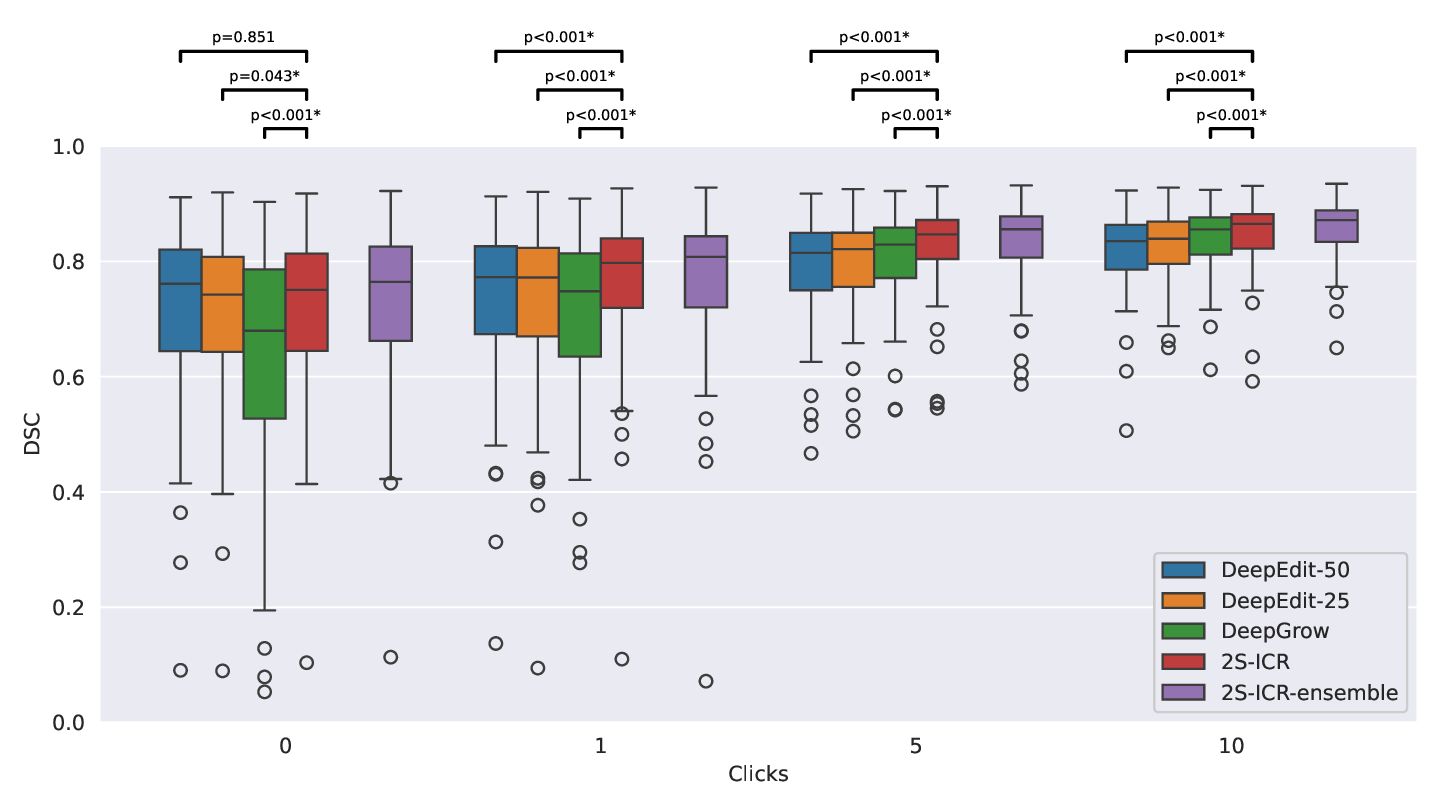}
  \caption{Model comparison on segmentation performance change in Dice similarity coefficient (DSC) evaluated on the MDA dataset. The statistical significance tests between the models are based on the two-sided Wilcoxon signed rank test with Benjamini–Hochberg procedure to correct for multiple testing, in which p $<$ 0.05 is considered significant.}
  \label{fig:overall_model_comparison_dsc}
\end{figure*}

Overall the 2S-ICR framework outperforms DeepGrow, DeepEdit-25, and DeepEdit-50 in the interactive segmentation performance. In addition, the 2S-ICR-ensemble provided additional performance outperforming the single model variant. The full comparison is shown in~\cref{fig:overall_model_comparison_dsc}.

As for the results on the MDA dataset, illustrated in~\cref{tab:mda-performance}, the 2S-ICR turned out to have the best performance. Indeed, the average 1 to 10 click DSC value was $0.819\pm0.106$ for 2S-ICR, $0.798\pm0.108$ for DeepGrow, $0.791\pm0.104$ for DeepEdit-25, and $0.785\pm0.109$ for DeepEdit-50. Furthermore, the ensemble version of 2S-ICR further amplifies these results, achieving the highest scores in each category. Specifically after ten clicks, the ensemble configuration reached the overall highest MDA dataset DSC value of $0.857\pm0.051$. 

\begin{table*}[h]
\centering
\resizebox{\textwidth}{!}{%
\begin{tabular}{c|c|c|c|c|c}
\hline
Model & 0 Clicks & 1 Click & 5 Clicks & 10 Clicks & 1 to 10 Avg. \\ \hline
2S-ICR (ours) &\textbf{ 0.713} ± \textbf{0.152} & \textbf{0.762} ± \textbf{0.136} &\textbf{ 0.824} ± \textbf{0.099} & \textbf{0.847} ± \textbf{0.079} & \textbf{0.819} ± \textbf{0.106} \\ \hline
DeepEdit-0 (DeepGrow)~\cite{deepgrow_2019, DeepEdit_2022} & 0.624 ± 0.226 & 0.707 ± 0.156 & 0.807 ± 0.089 & 0.839 ± 0.069 & 0.798 ± 0.108 \\ \hline
DeepEdit-25~\cite{DeepEdit_2022} & 0.707 ± 0.157 & 0.727 ± 0.148 & 0.794 ± 0.094 & 0.828 ± 0.064 & 0.791 ± 0.104 \\ \hline
DeepEdit-50~\cite{DeepEdit_2022} & 0.712 ± 0.157 & 0.733 ± 0.146 & 0.787 ± 0.101 & 0.819 ± 0.079 & 0.785 ± 0.109 \\ \hline
\hline
2S-ICR-ensemble (ours) & \textbf{0.722 ± 0.142} & \textbf{0.767 ± 0.131} & \textbf{0.832 ± 0.076} & \textbf{0.857 ± 0.051} & \textbf{0.827 ± 0.089} \\ \hline
\end{tabular}
}
\caption{The effect of interaction on the Dice similarity coefficient with the MDA dataset. Each interaction and sample outcome is averaged on three repetitions and the dataset (N=67) mean and standard deviation are reported. The best single model score of each category is bolded. Additionally, if ensemble has better performance, it is bolded.}
\label{tab:mda-performance}
\end{table*}

In the interactive segmentation setting with the HECKTOR dataset, it turned out that all the models benefited from interactions, as shown in~\cref{tab:hecktor-performance}. Furthermore, the 2S-ICR shows the highest performance out of the models with the average 1 to 10 click DSC value of  $0.845\pm0.118$, in comparison to $0.833\pm0.120$ for DeepGrow, and $0.825\pm0.124$ for DeepEdit-25, and $0.822\pm0.134$ for DeepEdit-50. The highest DSC of all the models was found to be after 10 interactions, the values of which were $0.871\pm0.066$ for 2S-ICR, $0.861\pm0.073$ for DeepGrow, $0.852\pm0.075$ for DeepEdit-25, and $0.849\pm0.087$ for DeepEdit-50.

\begin{table*}[h]
\centering
\resizebox{\textwidth}{!}{%
\begin{tabular}{c|c|c|c|c|c}
\hline
Model & 0 Clicks & 1 Click & 5 Clicks & 10 Clicks & 1 to 10 Avg. \\ \hline
2S-ICR (ours) & \textbf{0.752} ± \textbf{0.203} & \textbf{0.789} ± \textbf{0.190} & \textbf{0.851} ± \textbf{0.102} & \textbf{0.871} ± \textbf{0.066} & \textbf{0.845} ± \textbf{0.118} \\ \hline
DeepEdit-0 (DeepGrow)~\cite{deepgrow_2019, DeepEdit_2022} & 0.663 ± 0.316 & 0.757 ± 0.210 & 0.841 ± 0.099 & 0.861 ± 0.073 & 0.833 ± 0.120 \\ \hline
DeepEdit-25~\cite{DeepEdit_2022} & 0.729 ± 0.243 & 0.768 ± 0.196 & 0.829 ± 0.112 & 0.852 ± 0.075 & 0.825 ± 0.124 \\ \hline
DeepEdit-50~\cite{DeepEdit_2022} & 0.738 ± 0.229 & 0.768 ± 0.197 & 0.824 ± 0.123 & 0.849 ± 0.087 & 0.822 ± 0.134 \\ \hline
\end{tabular}
}
\caption{The effect of interaction on the Dice similarity coefficient with the HECKTOR dataset. Each interaction and sample outcome is averaged on three repetitions and the dataset (N=224) mean and standard deviation are reported. The best score of each category is bolded.}
\label{tab:hecktor-performance} 
\end{table*}

\subsection{Non-interactive segmentation performance}

The non-interactive MDA dataset results, presented in \cref{tab:mda-performance}, show that 2S-ICR achieves the highest performance out of the models with a DSC value of $0.713\pm0.152$. Consistent with the observations from the HECKTOR dataset, DeepGrow exhibited the lowest performance in this setting, with a DSC of $0.624\pm0.226$. The DeepEdit models showed improved performance, achieving DSC values of $0.707\pm0.157$ and $0.712\pm0.157$ for 25\% and 50\% click-free iterations, respectively. The ensemble version of 2S-ICR further enhanced these results, reaching a DSC value of $0.722\pm0.142$ in this non-interactive setting.

As for the HECKTOR dataset results, presented in \cref{tab:hecktor-performance}, when comparing 2S-ICR to DeepGrow and DeepEdit without interactions, it demonstrated the highest baseline performance, achieving a Dice similarity coefficient (DSC) of 0.752 $\pm$ 0.203. Our findings for DeepGrow and DeepEdit align with those reported in~\cite{DeepEdit_2022}, where DeepGrow exhibited the lowest performance without interactions, with a DSC of 0.663 $\pm$ 0.316. The DeepEdit approach turned out to improve upon DeepGrow, achieving DSC values of 0.729 $\pm$ 0.243 and 0.738 $\pm$ 0.229 for 25\% and 50\% click-free iterations, respectively. 

\begin{figure}[h!]
  \centering
   \includegraphics[width=\linewidth]{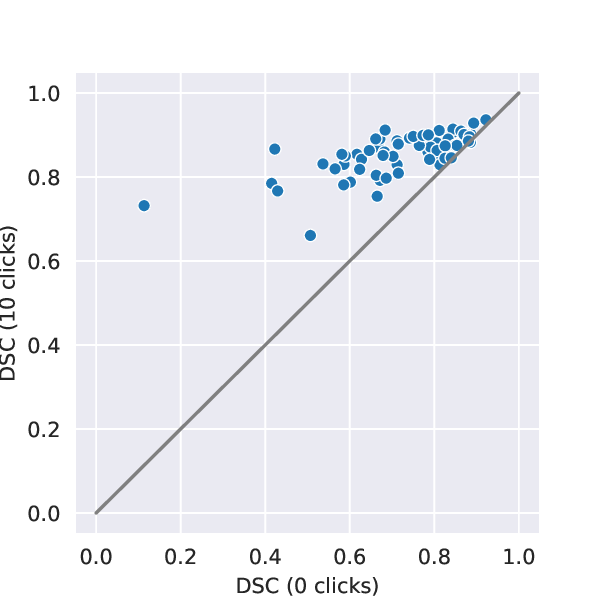}
  \caption{Segmentation performance change in Dice similarity coefficient (DSC) on individual samples from the MDA dataset using the ensemble version of 2S-ICR. The DSC with and without interactions is shown in y-axis and x-axis, respectively.}
  \label{fig:dsc_pre_after}
\end{figure}

In Figure~\ref{fig:dsc_pre_after}, improvements in segmentation are observed across all instances after interaction. Noteworthy enhancements are particularly evident in samples with initially suboptimal segmentations. Importantly, all samples demonstrated either consistent or improved segmentation accuracy, with no cases exhibiting a decline in quality due to the interaction process.

Segmentation refinement by the 2S-ICR framework is demonstrated using a scan from the MDA test set, where the network initially erroneously segmented two false-positive regions. The false positive region on the left side is removed with one click, and the false positive on the right side is removed with two clicks, totaling three click interactions for correction. This illustrates the method's capability to efficiently reduce false positives through user interactions. The initial segmentation and changes after first three interactions in false positives is illustrated in~\cref{fig:example_image}.

\begin{table*}[h]
\centering
\resizebox{\textwidth}{!}{%
\begin{tabular}{cccc}
\includegraphics[width=0.25\textwidth]{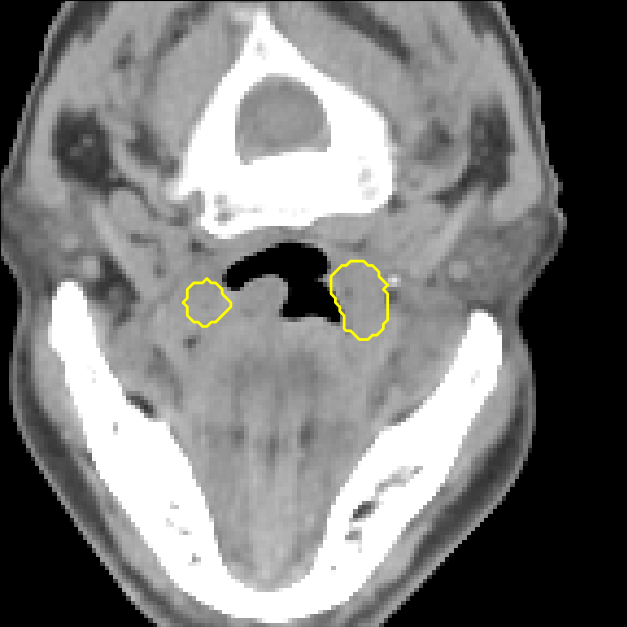} &
\includegraphics[width=0.25\textwidth]{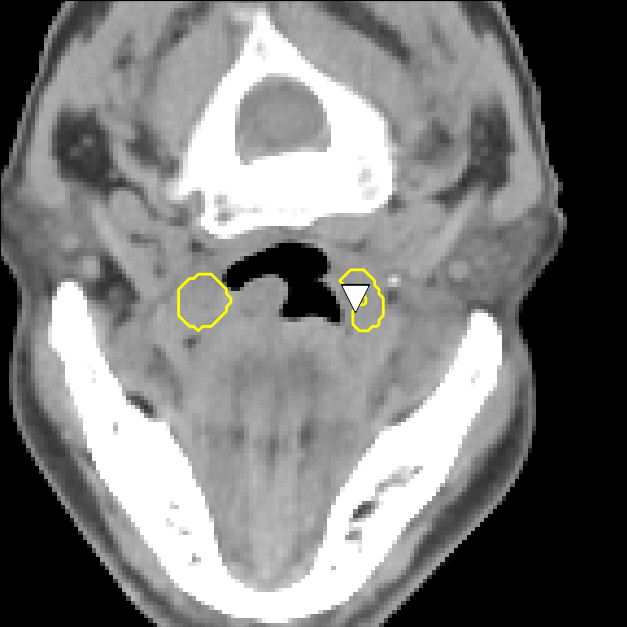} &
\includegraphics[width=0.25\textwidth]{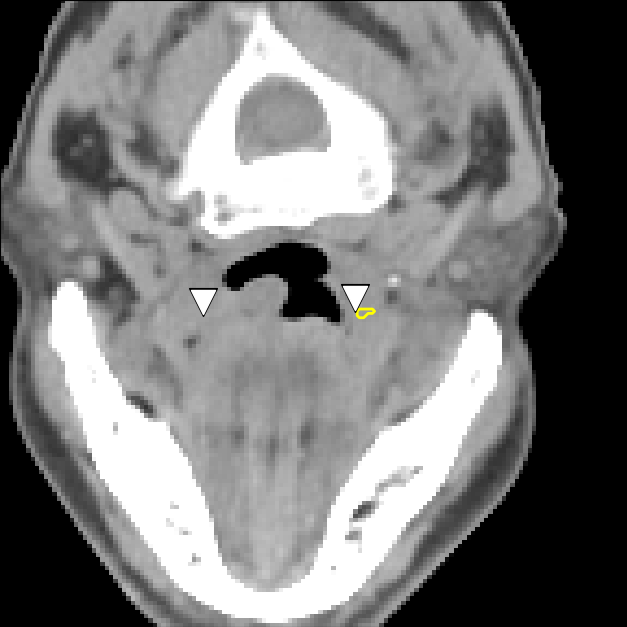} &
\includegraphics[width=0.25\textwidth]{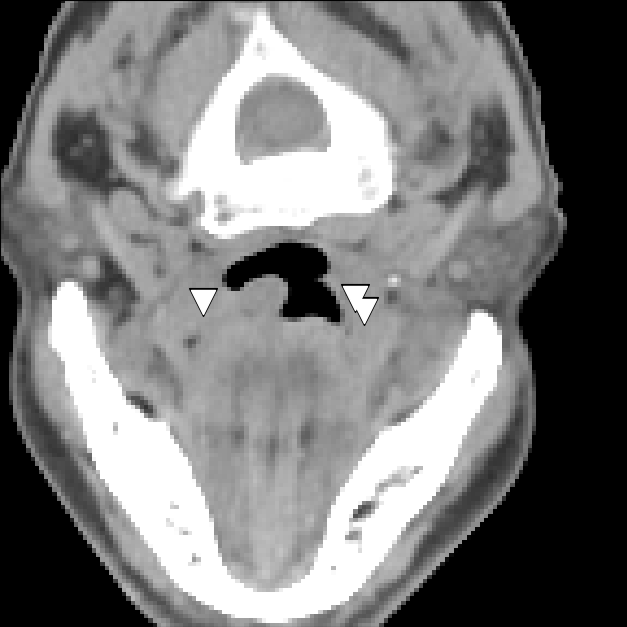} \\
\includegraphics[width=0.25\textwidth]{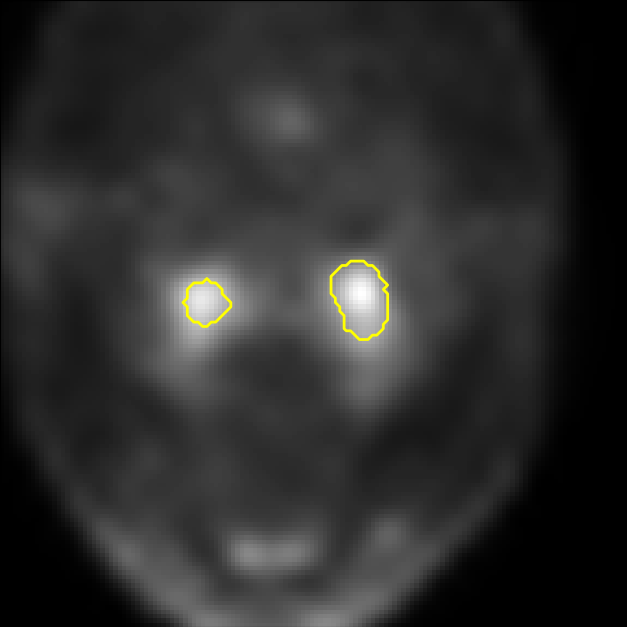} &
\includegraphics[width=0.25\textwidth]{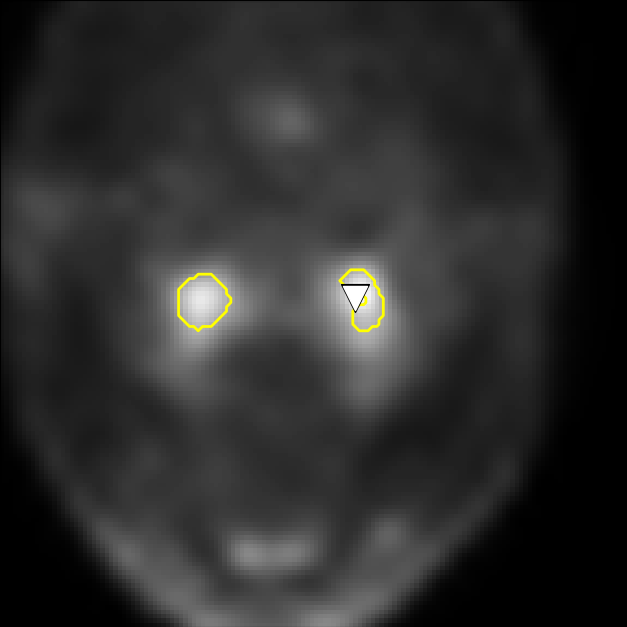} &
\includegraphics[width=0.25\textwidth]{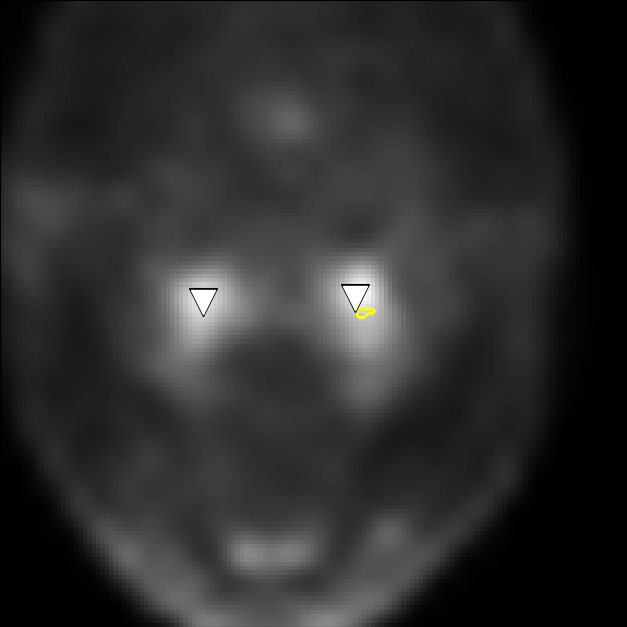} &
\includegraphics[width=0.25\textwidth]{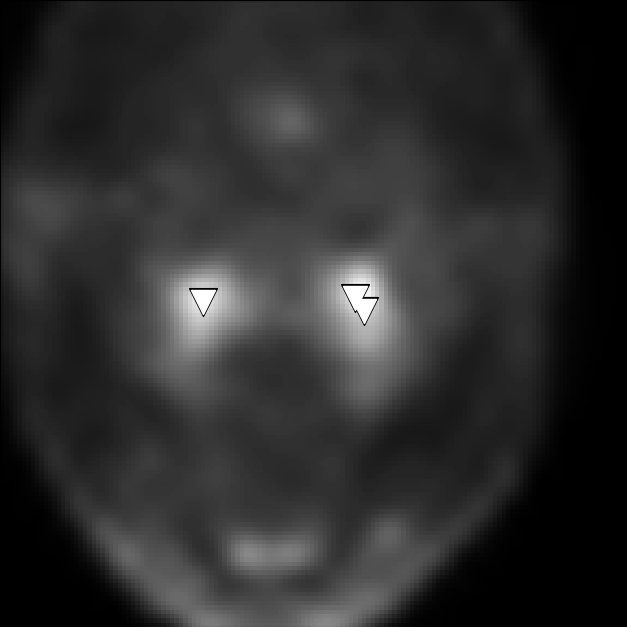} \\
\multicolumn{1}{c}{Baseline Segmentation} & 
\multicolumn{1}{c}{After 1 Interaction} & 
\multicolumn{1}{c}{After 2 Interactions} & 
\multicolumn{1}{c}{After 3 Interactions}
\end{tabular}
}
\captionof{figure}{Progressive refinement of segmentation through first three user interactions overlaid on CT (top row) and PET (bottom row) slices. False positives are marked in yellow and clicks with white arrow.}
\label{fig:example_image}
\end{table*}


For DeepEdit and DeepGrow, the results are consistent with the previous research on different medical domains~\cite{DeepEdit_2022}. Initially, the DeepEdit outperformed the DeepGrow without user interactions. However, this dynamic shifted with the introduction of user inputs. After five interactions, the DeepGrow began to surpass the DeepEdit, and this advantage increased with additional interactions. This outcome highlights the fundamental trade-off between non-interactive and interactive performance of the various implementations of the DeepEdit methodology.

\begin{table}[h]
\centering
\begin{tabular}{c|c|c}
\hline
$p$ & DSC 1 to 10 Avg. & N changed voxels  \\
\hline
0.0 & $0.827 \pm 0.134$ & $731 \pm 719$ \\
0.2 & $0.845 \pm 0.109$ & $941 \pm 1477$ \\
0.4 & $0.843 \pm 0.112$ & $888 \pm 1075$ \\
0.6 & $0.843 \pm 0.116$ & $932 \pm 1161$ \\
0.8 & $0.842 \pm 0.104$ & $917 \pm 1208$ \\
\hline
\end{tabular}
\caption{Effect of varying Mask Dropout probability (\( p \)) on interactive segmentation performance and the number of voxels changed by interactions. The table shows the average Dice similarity coefficient (DSC) and the number of changed voxels per interaction event (\( \mu \pm \sigma \)).}
\label{tab:click_comparison}
\end{table}

The introduction of mask dropout in 2S-ICR influenced significantly the interactive segmentation performance. Although the average Dice similarity coefficient (DSC) showed marginal fluctuation, the number of voxels altered per interaction event varied notably with different dropout probabilities. Specifically, when \( p = 0 \), the changes per interaction were substantially smaller compared to scenarios where \( p > 0 \). Additionally, the increased standard deviation of the changed voxels for \( p > 0 \) suggests that the model is capable of making more substantial adjustments when necessary. Note that while significant differences are observed between \( p = 0 \) and \( p > 0 \), variations for non-zero \( p \) values are minimal, indicating that the presence of any mask dropout ($p>0$) is more critical than its exact value.


\section{Discussion}

This study introduces 2S-ICR framework, a novel interactive deep learning method that sets a new standard for segmenting primary gross tumor volume in oropharyngeal cancer. Evaluated on the HECKTOR and MD Anderson Cancer Center datasets, the proposed framework consistently outperforms the previous state-of-the-art methods, displaying its ability to leverage user interactions without sacrificing baseline performance.

Volumetric medical image segmentation holds immense potential, but also presents unique challenges inherent to the medical domain, such as heterogeneous data from various imaging devices, imaging artifacts, and patient-specific variations. These complexities can lead to occasional failures in AI-driven segmentation, underscoring the critical need for human expertise to guide and refine the process interactively with AI models.

While the interactive segmentation has strong background in 2D, particularly in non-medical domains~\cite{liu2023simpleclick, sun2024cfr, sofiiuk_revieving_2022_ICIP}, the previous interactive segmentation research on the OPC GTVp domain has primarily focused on 2D slicing methods~\cite{wei2023opc_slice_idl} or reducing the annotation effort~\cite{luan2024semi_opc}. However, the state-of-the-art non-interactive segmentation methods for OPC GTVp have utilised 3D methods with the volumetric PET-CT scans and was found to improve the performance via global context in comparison to 2D methods~\cite{andrearczyk_hecktor2021_overview, andrearczyk2023_hecktor2022_summary, myronenko_hecktor_winner_2022, sahlsten2024hecktor, Iantsen_2021_hecktor_2020_winner}. Our work addresses this limitation by performing interactive OPC GTVp segmentation directly in volumetric space.

While some 2D interactive segmentation methods employ two models to reduce computational costs~\cite{chen2022focalclick_CVPR}, our motivation for DeepRefineNet's two-model architecture is distinct. We prioritize avoiding the trade-off between non-interactive and interactive performance often seen in single-model approaches~\cite{DeepEdit_2022}. By leveraging previous outputs as input, a common practice in 2D shown to stabilize predictions~\cite{sofiiuk_revieving_2022_ICIP}, we not only enhance 3D performance but also seamlessly chain non-interactive and interactive models, which enables a synergistic workflow where each model is optimized for its specific task.

DeepEdit was one of the first interactive models implemented for 3D medical segmentation tasks~\cite{DeepEdit_2022}, where both the pre- and post-interaction performances were measured. In DeepEdit, it was found that the quality of interactive DL segmentation without interactions was worse than that of a non-interactive DL model. DeepEdit has addressed this issue, to some extent, with the approach of "click-free" (i.e., non-interactive) training iterations. However, this approach introduced a trade-off: more click-free iterations improved non-interactive performance at the expense of interactive performance. In contrast, DeepRefineNet's two distinct models ensure optimal initial segmentation and effective refinement with interactions.

The analysis of both the HECKTOR and MDA datasets reveals significant variance in image level segmentation results. This observation aligns with the existing literature, highlighting the challenges of accurate GTVp segmentation~\cite{rasch2005target, lin2023pluribus}. This variability underscores that while some segmentations are clinically adequate, others fall short, which highlights the need for an interactive method to effectively improve sub-optimal segmentations.

Our study has a few limitations. First, the 2S-ICR framework has only been developed and evaluated with a binary segmentation task while clinical practice often demands multi-class segmentation, such as differentiating between primary tumors and lymph nodes~\cite{andrearczyk2023_hecktor2022_summary}. However, we note that our approach is not limited to binary segmentation tasks. Indeed, these limitation can be addressed in the future with datasets that encompass diverse medical tasks, image modalities (e.g., MRI), and the challenges of interactive multi-class segmentation. Second, we have here employed simulated interaction events, following the established algorithm in~\cite{deepgrow_2019, DeepEdit_2022}. While simulations provide a controlled environment, they may not fully describe how clinicians perform the interaction. Indeed, the algorithm uses a heuristic to select interaction coordinates by favouring areas with large errors and never provides an incorrect label. Furthermore, our preliminary results indicated that the interaction location had a considerable effect on the model performance that highlights the need to understand clinician behavior during interactive segmentation for improved applicability. In order to develop better suited interaction simulation algorithms, human interaction patterns should be analysed, which is left for future work.


The benefits of the proposed framework extend beyond improved accuracy. By eliminating the last remaining drawback associated with interactive segmentation, 2S-ICR unlocks the full potential of the entire field. This breakthrough paves the way for wider adoption of interactive segmentation across various clinical applications. By enabling clinicians to easily and quickly improve segmentation results, it promises more accurate treatment planning and improved patient outcomes.

In conclusion, this study introduces 2S-ICR, a novel click-based interactive framework, for the segmentation of primary gross tumor volume in oropharyngeal cancer. Our results show that the framework achieves comparable or superior performance to state-of-the-art interactive deep learning methods, both with and without user interactions. These results highlight the potential of the approach to enhance GTVp segmentation, by enabling clinicians to quickly improve segmentation results based on just a few interactions. The more accurate segmentation enabled by our approach could lead to more precise OPC treatment planning and improved patient outcomes.


\section{Methods}

\begin{figure*}[ht]
  \centering
  \includegraphics[width=\textwidth]{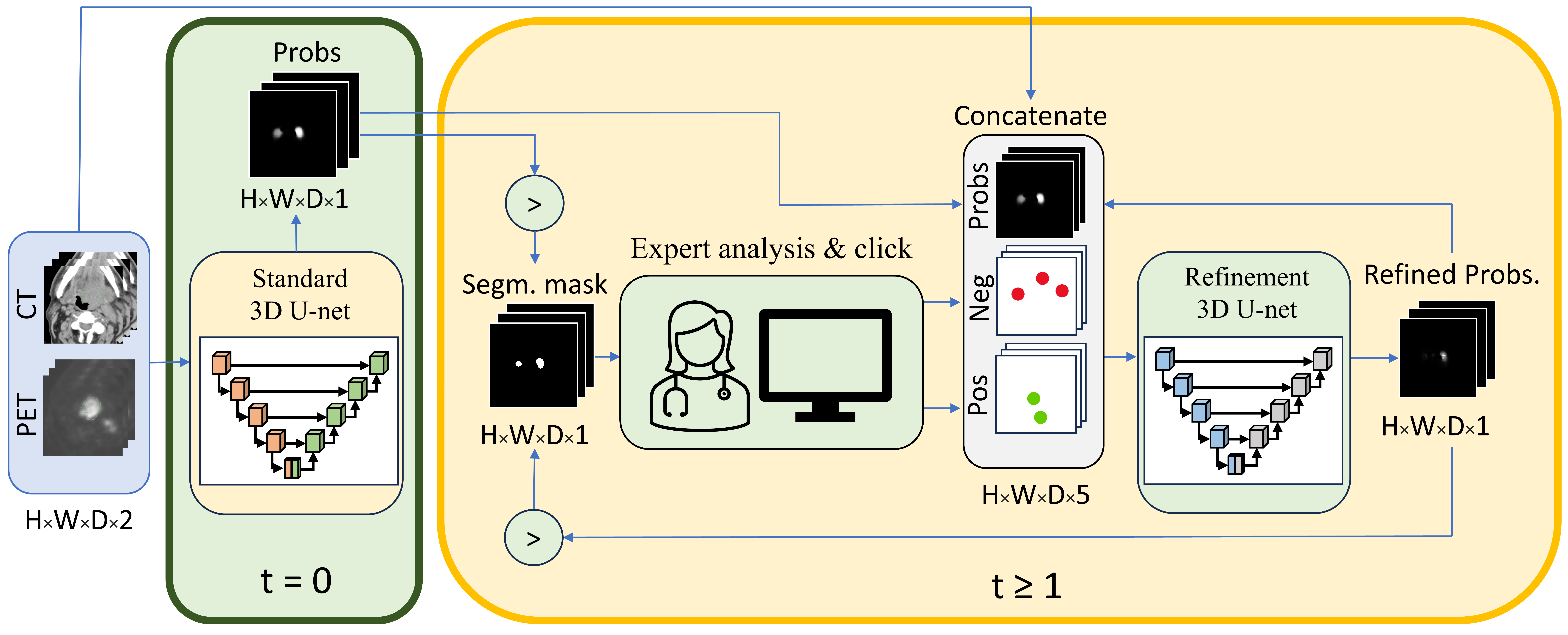}
    \caption{Visualisation of 2S-ICR framework. The initial segmentation ($t=0$) is provided by a standard model which is shown in the green box on left. The segmentation refinement ($t\geq1$) loop using a refinement model is visualised in the yellow box on right. Spatial dimensions (H$\times$W$\times$D), thresholding ($>$), negative (Neg), and positive (Pos) feature maps.}
  \label{fig:graphical_abstract}
\end{figure*}


In this section, we provide a detailed overview of the datasets, models, and evaluation metrics used to produce the results. Our proposed dual-model segmentation framework is illustrated in \cref{fig:graphical_abstract}. This study does not involve human subjects as it relies on retrospective and registry-based data; therefore, it is not subject to IRB approval. Our external validation dataset was retrospectively collected under a HIPAA-compliant protocol approved by the MD Anderson Institutional Review Board (RCR03-0800), which includes a waiver of informed consent. 

\subsection{Datasets}


The 2021 HEad and neCK TumOR dataset (HECKTOR), introduced in~\cite{andrearczyk_hecktor2021_overview}, consists of co-registered PET-CT images from 224 patients. The dataset was gathered from five centers located in Canada, Switzerland, or, France, with ground truth GTVp segmentations provided by multiple annotator agreement. The external MD Anderson Cancer Center dataset (MDA) consists of co-registered PET-CT images from 67 patients that are human papillomavirus positive, with the GTVp segmentations from a single annotator. The images are cropped to contain only the head and neck region, centered on the GTVp, and resampled to $144^3$ volumes with isotropic 1~mm resolution, i.e., in terms of both pixel-spacing and slice thickness.

We adhered to the data normalization procedure established in the previous work~\cite{sahlsten2024hecktor}. The CT scans were windowed to [-200, 200] Hounsfield units and subsequently normalized to the range of [-1, 1]. The PET scans were standardized using z-score normalization. This normalization procedure ensured consistency across the datasets and enabled the use of the same models without retraining.

\subsection{2S-ICR framework}

The 2S-ICR framework, as described in \cref{alg:dualrefinenet}, introduces a novel approach to interactive PET-CT segmentation by combining a standard deep learning (DL) model and an interactive DL model.

The iterative refinement algorithm of the 2S-ICR framework is as follows:

\begin{algorithm}[!h]
\caption{An algorithm with caption}\label{alg:dualrefinenet}
\begin{algorithmic}
\Require DL model $f_{DL}$, interactive DL model $f_{IDL}$, PET-CT scan $x$, user interaction coordinate model $g$
\State $C = \emptyset$
\State $\hat{y} = f_{DL}(x)$
\State $c = g(\hat{y})$
\While{$c \neq \emptyset$}

$C = C \cup c$

$\hat{y} = f_{DL}(x, \hat{y}, C)$

$c = g(\hat{y})$
\EndWhile
\end{algorithmic}
\end{algorithm}

\begin{align*}
    \hat{y}^{(t)} &=
    \begin{cases} 
        f_{DL}(x) & \text{if t = 0} \\
        f_{IDL}(x, \hat{y}^{(t-1)}, C^{(t-1)}) & \text{if } t \geq 1
    \end{cases}\\
    C^{(t)} &= 
    \begin{cases} 
        \emptyset & \text{if t = 0} \\
        C^{(t-1)} \cup \{c(\hat{y}^{(t)})\}\  & \text{if } t \geq 1        
    \end{cases}
\end{align*}

where $f_{DL}$ is the standard DL segmentation model, $f_{IDL}$ is the interactive DL model, $t$ is a non-negative integer denoting how many interaction events have occurred, and $C^{(t)}$ is the set of all interactions encountered at interaction event $t$. When t equals 0, it represents the initial segmentation, which is subsequently refined through user interactions or simulated interactions.

The principle behind our novel 2S-ICR framework is presented in \cref{alg:dualrefinenet}. It consists of two deep learning models, namely a standard DL segmentation model and an interactive DL segmentation refinement model. If no interactions are given to the 2S-ICR, it segments the PET-CT image using the standard model. When the user first interacts with the model, the given error coordinate, PET-CT image, and the output of the standard model are given to the interactive model. When the user further interacts with the 2S-ICR, the output of the standard model is replaced with the last output of the 2S-ICR, and the new interaction coordinate is given alongside with the previous ones for the 2S-ICR to further refine its output. We train the standard model and the interactive model separately, so as to closely follow the scenario where the 2S-ICR is applied on top of a pre-trained GTVp segmentation network.


As large-scale training and validation of an interactive DL model is not feasible with human interactions, we chose to simulate interactions in these phases. We used the user click interaction simulator proposed in~\cite{deepgrow_2019, DeepEdit_2022}. The click simulator compares the model output to the ground truth segmentation in order to select optimal interaction coordinates. Specifically, the simulator first extracts erroneous regions by examining where in the volume the model output and the ground truth differ. Then for each erroneous voxel, the distance to the border of the erroneous region is computed. After this, the distances are normalized by the sum of all the distances. As a result, all voxel values are in the range [0, 1] and sum to 1. Then, we treat these values as probabilities of a multinomial distribution and sample an interaction coordinate accordingly.

We utilized the Monai implementation of the 3D U-Net architecture~\cite{3dunet} across all models: the initial segmentation model of 2S-ICR framework, the segmentation refinement model of the framework, and the models for DeepGrow and DeepEdit. To ensure a fair comparison between these methods, the only difference was the number of input channels The networks consisted of channels [16, 32, 64, 128, 256], stride [1, 2, 2, 2], two residual units. The choice of a stride of 1 for the first layer was pivotal for enhancing the impact of the interaction event.

The 2S-ICR-ensemble experiments consist of a five-member ensemble based on the 5-fold cross-validation, both for the standard and interaction models. The ensemble probability is based on the average of the member probabilities. The ensembling was integrated into the interaction loop by selecting each interaction coordinate based on the ensemble prediction, i.e., each member in interaction model is given the same interaction coordinates in each iteration.

\subsection{Training Procedure}


The refinement network of 2S-ICR can become overly reliant on the previous model output mask, which it uses as one of its inputs. This phenomenon hinders its ability to utilise interaction feedback effectively. To address this, we developed a novel regularization approach used in the segmentation refinement model training. The approach randomly omitted the previous mask ($p=0.2$) and replaced it with a volume filled with the value 0.5. As the previous model output mask is the post-sigmoid output, i.e., each voxel represents the probability of the foreground class, we selected the value of 0.5 to signify no information regarding the class. This regularization encourages the network to rely more on the original input and user interactions, and less on the previous mask, which we observed to lead to improved performance and greater responsiveness to interaction coordinates. The impact of varying this omission probability on both the  segmentation performance and number of modified voxels during each interaction. The standard deviation of changed voxels increases significantly when ($p>0$), indicating more flexible voxel adjustments for each interaction event. Full results are presented in the Supplementary Table~\ref{tab:click_comparison}.

The refinement network of the 2S-ICR framework was optimized in every interaction event during training, unlike in case of previous interactive volumetric segmentation methods~\cite{DeepEdit_2022}. Indeed, the backward pass of earlier studies was executed only after all the interaction events have been accumulated, which is computationally costly. We chose to train the refinement network in this way, as we observed it to speed up considerably the training process, while still allowing the model to reach state-of-the-art performance.

Following the previous research, we randomly determine the number of simulated interactions in each training iteration~\cite{DeepEdit_2022}. While the previous research employed a uniform distribution ranging from 0 to 15, for the mask refinement component of the DeepRefineNet framework, we utilized a range of 1 to 15. During validation, a constant number of 10 interactions is maintained.

All the models were trained according to a composite loss function that integrates the Dice Loss with the binary Cross-Entropy Loss, similarly to previous approaches for the OPC GTVp segmentation in the literature~\cite{sahlsten2024hecktor, myronenko_hecktor_winner_2022}. The composite loss function we use is formulated as:
\begin{equation}
\mathcal{L}_{\text{DiceBCE}} = \mathcal{L}_{\text{Dice}} + \mathcal{L}_{\text{BCE}},
\end{equation}
where the Dice Loss is denoted with $\mathcal{L}_{\text{Dice}}$ and the binary Cross-Entropy loss with $\mathcal{L}_{\text{BCE}}$. While the composite loss may also be computed as a weighted sum of the components, we chose to use uniform weights.

The Dice Loss is defined by the following equation:
\begin{equation}
\mathcal{L}_{\text{Dice}} = 1 - \frac{2 \sum_{i} p_{i}g_{i} + \epsilon}{\sum_{i} p_{i} + \sum_{i} g_{i} + \epsilon},
\end{equation}
where $p$ and $g$ represent the model output and ground truth segmentation, respectively, and $\epsilon=1\times 10^{-5}$ is a smoothing factor to prevent division by zero. Furthermore, the Binary Cross-Entropy Loss is defined as:
\begin{equation}
\mathcal{L}_{\text{BCE}} = -\sum_{i} g_{i} \log(p_{i}) + (1-g_{i}) \log(1-p_{i}).
\end{equation}
In both the Dice Loss and the Binary Cross-Entropy Loss, the summation is over the voxels of the segmentation volume. The composite loss allows for a more comprehensive DL segmentation optimization by addressing both the overlap of imbalanced foreground class and the per-pixel classification accuracy~\cite{asgari2021deep_review}.

To enhance the models' robustness towards the variations in imaging conditions and to reduce overfitting, we apply a data augmentation pipeline during model training. The data augmentation approach adheres to the procedures presented in~\cite{myronenko_hecktor_winner_2022}. All the models are trained under the same settings for fair comparison. Specifically, we apply random affine transformations that include rotations over all axes up to 45 degrees, scaling and shearing within ranges of [-0.1, 0.1], and translation within a range of [-32, 32] voxels, each with a probability of 0.5. In addition, the augmentations include mirroring over all the axes, induced with the same probability of 0.5. For the CT modality, additional intensity augmentations are implemented, which consist of random contrast adjustments, with gamma range of [0.5, 1.5] and applied with probability of 0.25, intensity shifting, with offset of 0.1 and applied with probability of 0.25, random Gaussian noise, with standard deviation of 0.1 and applied with probability of 0.25, and Gaussian smoothing applied with probability of 0.25.

The training of all the models spanned at most 300 epochs with an early stopping patience set at 50. We utilized the AdamW~\cite{adamw} optimization algorithm to reduce overfitting. The initial learning rate was selected as $10^{-4}$ that gradually decayed to zero by the final epoch using the Cosine Annealing scheduler. The AdamW weight decay coefficient was set to $10^{-5}$. The best checkpoint of each model was based on the Dice coefficient, which was evaluated after each epoch on the 5-fold cross validation test fold. Each model was trained using a mini-batch size of 1. The models were trained on an NVIDIA A100 GPU with 80 GB of memory. The extra validation runs of the trained models were performed using an NVIDIA RTX 3080 GPU with 10 GB of memory.

\subsection{Evaluation Measures}

The segmentation performance is evaluated with the Dice Similarity Coefficient (DSC), average symmetric surface distance (ASSD), Hausdorff distance at 95\% (HD95), and Surface Dice Similarity Coefficient at 1~mm (SDSC). The DSC is a measure of the overlap between the predicted segmentation volume $\hat{y}$ and the ground truth segmentation $y$, defined as follows:
\begin{equation}
\text{DSC} = \frac{2 \sum_{i} \hat{y}_{i}y_{i}}{\sum_{i} \hat{y}_{i} + \sum_{i} y_{i}}.
\end{equation}
The definition is similar to that of the Dice Loss but it is calculated with the thresholded values of the output probability volume $\hat{y} = g > 0.5$. 

The ASSD metric measures the average distance between the boundaries of the predicted and ground truth segmentation volumes with following definition:
\begin{gather}
    \text{ASSD} = \frac{D}{|S(y)| + |S(\hat{y})|}, \\
    D = \sum_{p \in S(y)} \min_{q \in S(\hat{y})} d(p, q) + \sum_{q \in S(\hat{y})} \min_{p \in S(y)} d(q, p) \nonumber
\end{gather}
where $S(\cdot)$ is an operation that extracts the set of surface points and $d(\cdot,\cdot)$ is the Euclidean distance. The distance between a point and a set of points is the minimum distance between the point and any element of the set.

The HD95 metric measures the 95th percentile of the largest distance among the closest point pairs between the ground truth and model segmentations. It is defined as follows:
\begin{align}
    \text{HD95} = & \max \{ \max_{\text{P}95,~p \in S(\hat{y})} \min_{q \in S(y)} d(p, q), \\ \nonumber 
    & \max_{\text{P}95,~p \in S(\hat{y})} \min_{q \in S(y)} d(q, p) \}.
\end{align}

The SDSC metric measures the overlap on the segmentation boundaries at a specific threshold:
\begin{equation}
    \text{SDSC} = \frac{2 \times |S_\tau(y) \cap S_\tau(\hat{y})|}{|S_\tau(y)| + |S_\tau(\hat{y})|},
\end{equation}
where $S_\tau(\cdot)$ is an operation that extracts the set surface voxels with a margin $\tau$, which was selected as 1mm in our study.

\section{Data availability}

Data availability The HECKTOR 2021 training dataset~\cite{andrearczyk_hecktor2021_overview} is publicly accessible from \url{https://www.aicrowd.com/challenges/miccai-2021-hecktor}. The external validation dataset is publicly available on Figshare (\url{https://doi.org/10.6084/m9.figshare.22718008}) under CC BY 4.0 license.

\section{Code Availability}

The 2S-ICR framework is available on GitLab (Link will be added later). This repository includes comprehensive setup instructions for environment configuration and dependency management, facilitating straightforward installation and deployment. Links to trained models hosted on OneDrive are also provided.

\bibliographystyle{ieeetr}  
\bibliography{references}  

\end{document}